# ONTOLOGY FOR MOBILE PHONE OPERATING SYSTEMS

Hasni Neji and Ridha Bouallegue

Innov'COM Lab, Higher School of Communications of Tunis, Sup'Com
University of Carthage, Tunis, Tunisia.

*Email:* `hasni.neji63@laposte.net; ridha.bouallegue@gmail.com`

## ABSTRACT

*This ongoing study deals with an important part of a line of research that constitutes a challenging burden. It is an initial investigation into the development of a Holistic Framework for Cellular Communication (HFCC). The main purpose is to establish mechanisms by which existing wireless cellular communication components and models can work holistically together. It demonstrates that establishing a mathematical framework that allows existing cellular communication technologies (and tools supporting those technologies) to seamlessly interact is technically feasible.*

*The longer-term future goals are to actually improve the interoperability, the efficiency of mobile communication, calls quality, and reliability by applying the framework to specific development efforts.*

## 1. INTRODUCTION

Operating system's Interoperability is the ability of independent software systems (tailored to mobile phones) to work with each other, exchange information and initiate actions from each other, without the condition of knowing neither the details nor how they each work. However, to reach this objective, the domain of study must be explicitly understood without any ambiguities. The approach to this portion of investigation is to analyze the structures, and architectures of a small set of individual tools: Symbian Operating System (Symbian OS) -- a mobile operating system and computing platform designed for smart phones, Currently maintained by Accenture (is a global management consulting, technology services and outsourcing company headquartered in Dublin, Republic of Ireland)--. And Android Operating System (Android OS) -- a Linux-based operating system for mobile devices such as smart phones and tablet computers. The software is developed by the Open Handset Alliance led by Google)--. This includes performing a domain analysis (of this subset of tools) and building a feature model of the domain [1]. Next, the selected main artifact attributes will be considered in the context of the objects needed for establishing an object federation. Using this context, the characteristics will be defined within mobile phone's operating system Ontology [2]. This latter would be a pilot star-up framework for further extension by incorporating new mobile operating systems.

## 2. TOWARDS MOBILE PHONE OPERATING SYSTEM ONTOLOGY

### 2.1 Ontology overview:

The term "Ontology" is widely used in Knowledge Engineering and Artificial Intelligence where what "exists" are those entities which can be "represented." It is used to refer to the shared understanding of some domain of interest. It may be used as a unifying framework to solve problems in that domain [2].
Ontologies capture knowledge about some domain of interest. They describe the concepts in that domain and also the relationships that hold between those concepts.
To be effective, Ontology must entail or embodies a world view with respect to a given domain. This world view is often conceived as a set of concepts along with their definitions and their inter-relationships.

### 2.2 Methodology for building the Ontology

One way to overcome the drawbacks generated by the lack of interoperability in heterogeneous mobile phone's operating systems is to establish a unifying contextual and architectural framework for the domain. As this contextual and architectural framework or "Ontology" emerges; mobile phone's operating systems will be able to communicate with more efficiency.
In constructing the specific tool Ontologies (one for each mobile phone operating system); the focus is on identifying the classes (and methods) that were needed to pass objects from one tool to another.
Because there is currently no usable interoperability Ontology for the domain of mobile phone's operating systems, it was not possible to rely on previous works in designing a federation Ontology. Instead, this Ontology had to be constructed "from scratch."
Fortunately, there were existing methodologies for designing Ontologies. The methodology presented by [3] is one of them. It is possible to tailor this methodology to develop a specific Ontology for constructing the federation one. The Ontology development process consists of the following steps:

(1) identify the purpose and scope of the Ontology,
(2) perform a feature analysis for the domain of mobile phone operating systems' tools,
(3) collect similar characteristics between different feature models, establish affinity relationships, and group commonalities between the two tools in order to build a federation Ontology representing these commonalities and enter this Ontology into Protégé 4.1,
(4) construct the more detailed Ontologies for each tool in Protégé 4.1,
(5) use the Unified Modeling Language (UML) to represent the relationships between the three Ontologies, and
(6) Document the Ontologies.

**1. Step 1 -- Purpose and Scope of the Ontology**
In terms of purpose and scope, the federation Ontology must be broad enough to accommodate all potential mobile phone's operating systems, as well as being extensible in case new Ontology terms and relationships have to be added later. The specific development tool Ontologies must be detailed enough to account for the structures and architectures actually forming the basis of mobile phone's operating systems.
**2. Step 2 -- Feature Modeling**
The second step in the Ontology design methodology is to perform a domain analysis of mobile operating systems by constructing and then considering the feature models of Symbian OS and Android OS.

Feature modeling is a method used to help define software product lines and system families, to identify and manage commonalities and variabilities between products and systems [1].

Defining a feature model for an existing mobile phone's operating systems provides means to explore, identify, and define the key architectural aspects of the existing mobile phone's operating systems so that these aspects can then be described more fully in Ontology. It is this Ontology that can be used to establish interoperability between the existing mobile phone's operating systems.

As shown in figure 1, the feature model is defined around concepts and not around objects. The objective is to model features of elements and structures of a domain, not just objects in that domain. For more detail on the mechanism of how to construct a feature model, the reader may consult [1], [4], or [3]. In his book, Czarnecki [1] provides an excellent methodology for gathering the information needed to construct a feature tree. He identifies the sources of features as the following:

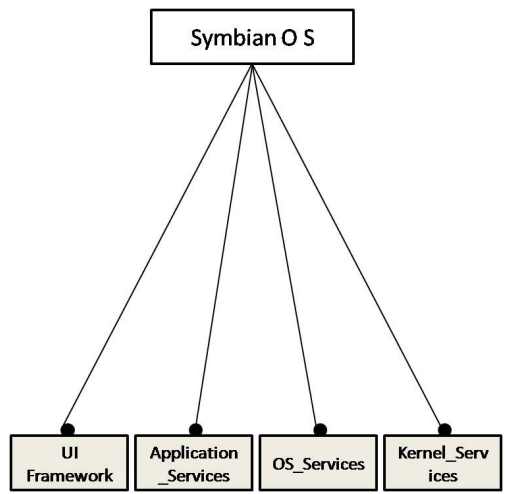

Figure 1. Feature Model of the Symbian high level concepts.

• Existing and potential stakeholders,
• Domain experts and domain literature,
• Existing systems,
• Pre-existing models (e.g., use-case models, object models…), and
• Models created during development process (i.e., features gotten during design).

Moreover, He identifies the following strategies for identifying and capturing features:
• Look for important domain terminology that implies variability.
• Use feature starter sets to start the analysis.
• Update feature models during the entire development cycle.
• Identify more features than you initially intend to implement.

Besides, according to [1] the following set of general steps illustrates the feature modeling process:
• Record similarities between instances (i.e. common features).
• Record differences between instances (i.e. variable features).
• Analyze feature combinations and interactions.
• Record all the additional information regarding features.
• Organize the features in hierarchical feature tree with classification (mandatory, optional, alternative, and/or optional alternative features).

International Journal of Wireless & Mobile Networks (IJWMN) Vol. 4, No. 3, June 2012

- ✓ Mandatory. A mandatory child feature must be included in all the products in which its parent feature is included.
- ✓ Optional. An optional child feature can be optionally included in all products in which its parent feature appears.
- ✓ Alternative. A set of child features are defined as alternative, if only one of them can be selected when its parent feature is part of the product. As an example, a mobile phone may use a Symbian or Android operating system but not both at the same time.

Or-relation. A set of child features are said to have an or-relation with their parent when one or more of them can be included in the products in which its parent feature appears. For instance, a mobile phone can provide connectivity support for Bluetooth, Universal Serial Base (USB), Wireless Fidelity (Wi-Fi) or any combination of the three.

### 3. Step 3 – Establishing Commonalities

After producing a feature model for Symbian and Android, the next step required is to isolate and annotate the commonalities (or the inferred commonalities) that exist between the two feature models. These common features then formed the basis for the basic Ontology terminology of the mobile phone's operating systems federation. The approach in this step is to brainstorm and reason about the two feature diagrams, develop lists of potential terms from the feature diagrams, identify common terms between the two lists, and then construct affinity diagrams of these common terms. Affinity diagrams are hierarchical Venn diagrams that provide groupings of related terms. Figure 2 illustrates how an affinity diagram is constructed; in this case the related terms all deal with "Application_Framework" as a part of the Android Operating System architecture.

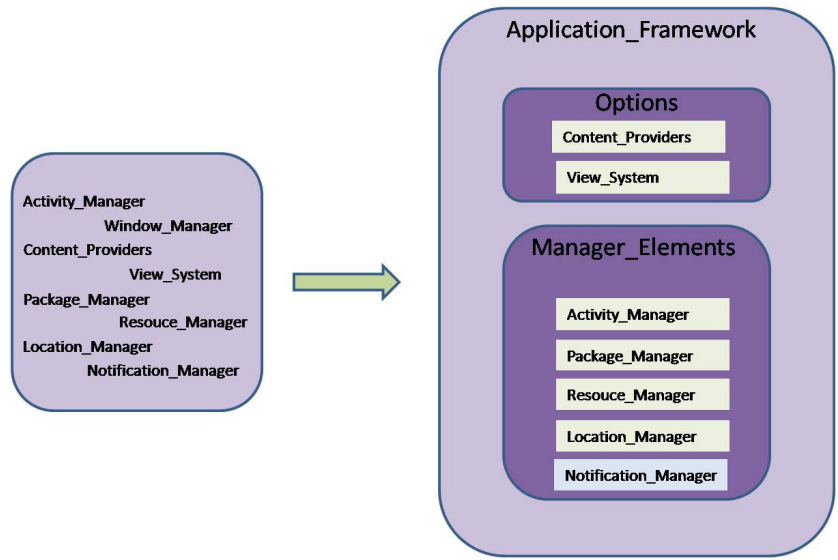

Figure 2.   Construction of an Affinity Diagram.

The groupings of terms in the affinity diagrams then provided the basis for the hierarchy of terms in the mobile phone's operating systems federation Ontology. These terms and their hierarchy were then entered and stored in the Protégé 4.1 ontology Web Language[1] (OWL).

### 4. Step 4 – Tool Ontologies

---
[1] http://protege.stanford.edu

As the terms of each Ontology's tool were identified they were input into the Protégé 4.1 Ontology Web Language. Different Ontology languages provide different facilities and fulfill the requirements needed for this work. The most recent development in standard Ontology languages is OWL from the World Wide Web Consortium (W3C). OWL makes it possible to describe concepts. It has a richer set of operators - e.g. intersection, union and negation. It is based on a different logical model which makes it possible for concepts to be defined as well as described. Complex concepts can therefore be built up in definitions out of simpler concepts. Furthermore, the logical model allows the use of a reasoner which can check whether or not all of the statements and definitions in the Ontology are mutually consistent and can also recognize which concepts under which definitions. The reasoner can therefore help to maintain the hierarchy correctly. This is particularly useful when dealing with cases where classes can have more than one parent.

OWL classes are interpreted as sets that contain individuals. They are described using formal (mathematical) descriptions that state precisely the requirements for membership of the class. One of the main services offered by a reasoner is to test whether or not one class is a subclass of another class. By performing such tests on the classes in Ontology it is possible for a reasoner to compute the inferred Ontology class hierarchy. Another standard service that is offered by reasoners is consistency checking. Based on the description (conditions) of a class the reasoner can check whether or not it is possible for the class to have any instances. A class is deemed to be inconsistent if it cannot possibly have any instances. In Protégé 4.1, it is possible to manually construct class hierarchy called the asserted hierarchy. The class hierarchy that is automatically computed by the reasoner is called the inferred hierarchy. Moreover, this tool automatically classifies the ontology (and check for inconsistencies).

The already mentioned reasons certify the high-quality of the tool. Protégé 4.1, OWL was chosen in order to make the descriptions of concepts, attributes, and instances formal, so as the knowledge can be machine-readable and reasoning-automated. Figure 3 is a screen capture of the protégé 4.1 OWL environment.

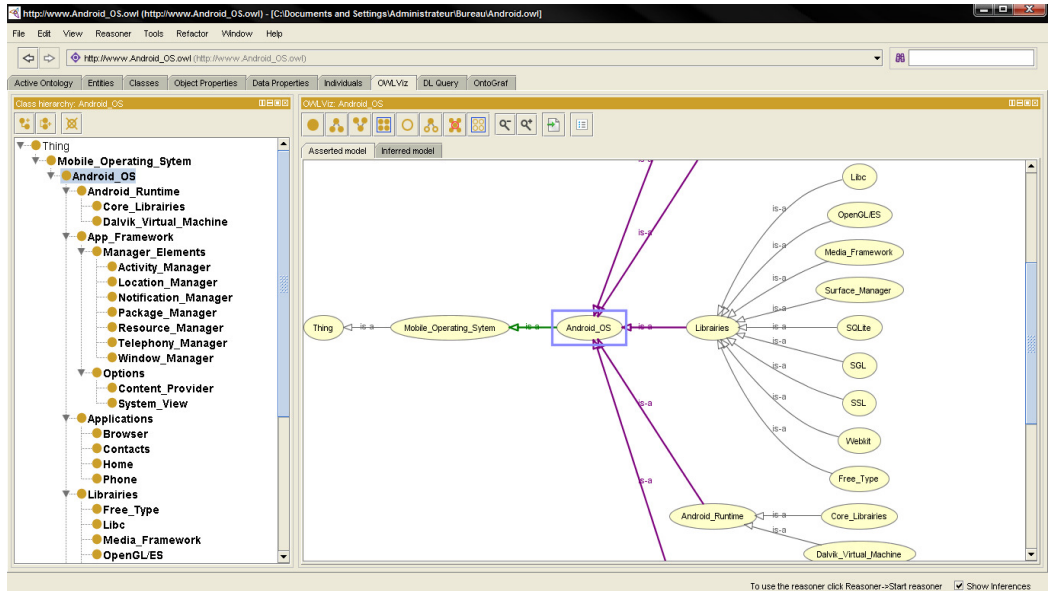

Figure 3. Screen shot of the protégé 4.1 tool.

## 5. Step 5 - UML Representation of the Domain

The relationships between all three Ontologies are identified and annotated. The reason for this is to formulate inter-relationships between the Ontologies.

Like Ontology, a well-defined class diagram, part of UML, can describe concepts and relationships in a certain domain.

Both approach have much to offer and can work together most effectively in an integrated environment. Figure 4 illustrates the general UML structure that was used to annotate how all three Ontologies were related.

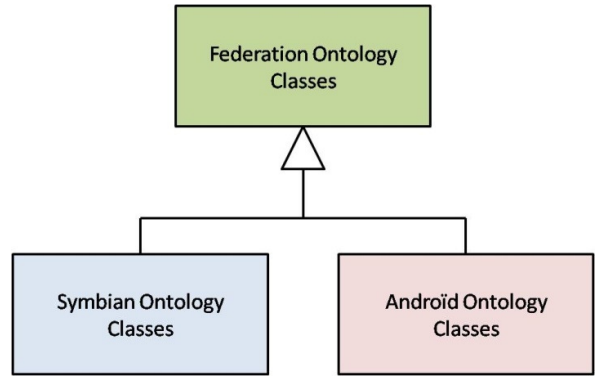

Figure 4**.** Ontology Inter-relationship.

## 6. Step 6 -- Documentation

The three developed Ontologies are self-documenting and are stored in Protégé 4.1 project files. Such documentation makes it possible for future researchers to modify the Ontologies or add additional tools to them. As a final note about the methodology used to develop these initial mobile phone's operating systems Ontologies, a modified version of this methodology should be used when adding additional tools to the HFCC federation. The modified methodology includes the following:

- Confirm that the purpose is still valid; expand the scope to include the new tool Ontologies. Remove invalid ones from the framework, they are no longer needed.
- Only perform feature modeling of the new tools so that needed constructs for the federation Ontology are identified. Since the federation Ontology is already established, it is only necessary to extend and modify it, not re-build it entirely.
- Modify the federation Ontology to account for the new/modified Ontology terms.
- Modify each UML relationship diagram as needed to account for the new tool Ontologies and the changes to the federation Ontology.

## 3. ONTOLOGY DESIGN

### 3.1 Domain analysis and feature models

A domain analysis of the mobile operating system's domain was undertaken. The analysis was accomplished by examining two specific mobile phone's operating systems, building feature models of those tools, and then identifying key terminology of the feature models. There are two reasons why this domain analysis cannot be considered to be a complete analysis of the domain of mobile phone's operating systems. First, only two tools (out of many of possibilities) were analyzed. Secondly, a domain analysis is not an "additive" activity; simply analyzing additional tools (beyond those two) by themselves does not completely add to the overall analysis. The ways in which the new additions affect and change the previously established analysis must also be considered. Therefore, the limited domain analysis conducted as part of this research can be considered a necessary, but not sufficient, analysis towards establishing a unified framework for mobile phone's operating systems.

### 3.1.1 Symbian Operating System

The first tool analyzed in the domain analysis was Symbian, a mobile operating system and computing platform designed for smart phones. The feature model was developed by identifying software features from the Symbian OS Architecture Source book [5] and by actual day-to-day use of the tool. Figure 5 illustrates a single excerpt from the overall feature model for Symbian. This excerpt illustrates the features associated with the architecture of Symbian operating system services, is just a small portion from the total architecture. Each of these features then becomes a candidate for possible inclusion in the federation Ontology. The complete list of Symbian features is shown below in figure 6. It is generated by Protégé 4.1 OWL tool.

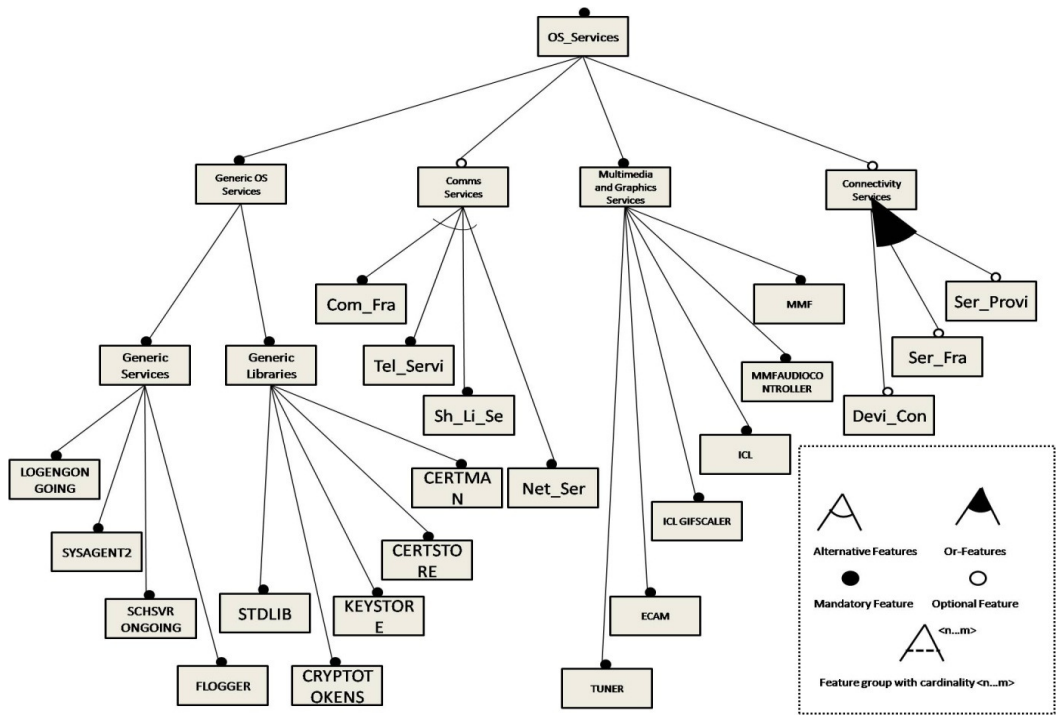

Figure 5. Operating System Services as an excerpt from the overall Symbian Feature Model.

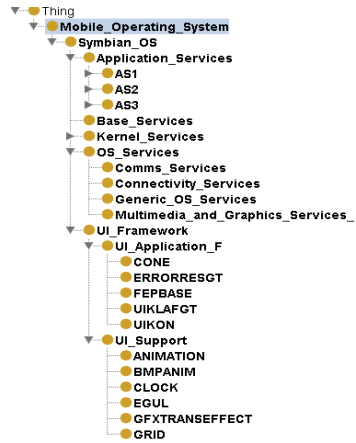

Figure 6. List of Symbian Operating System features.

This feature list is taken directly from the Symbian feature model. The features aligned as first level at the left of the figure are high-level "parent" features, while those second, third level,... etc represent more detailed "atomic" features.

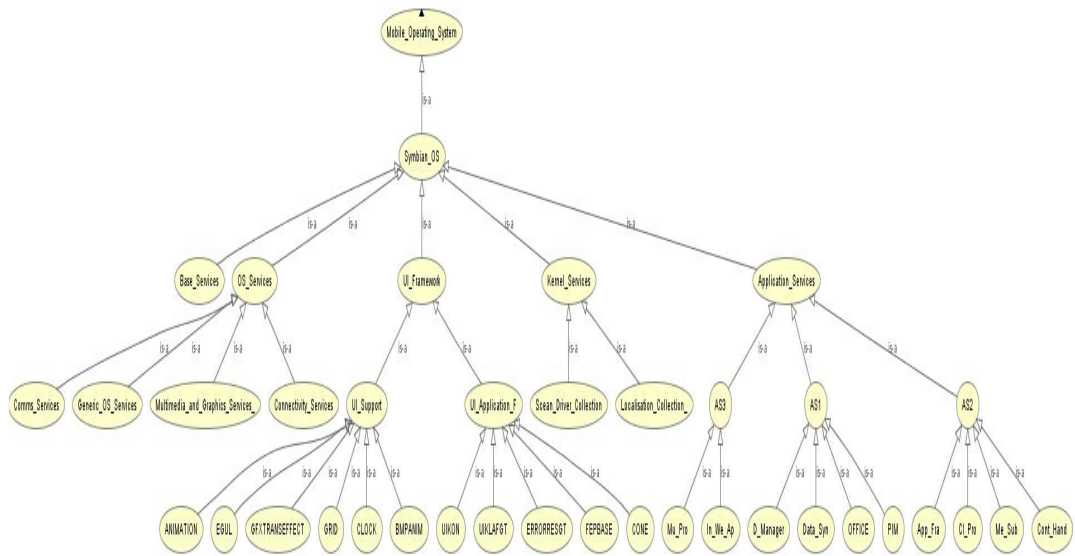

Figure 7. Symbian visualization ontology generated by the Protégé 4.1 tool.

Figure 7 is taken from the Ontology editor tool. It gives a thorough idea about the ontological concepts' hierarchy of Symbian operating system. Meanwhile, figure 8 presents the Ontological graph automated by OWL as layers of classes given by a top down approach.

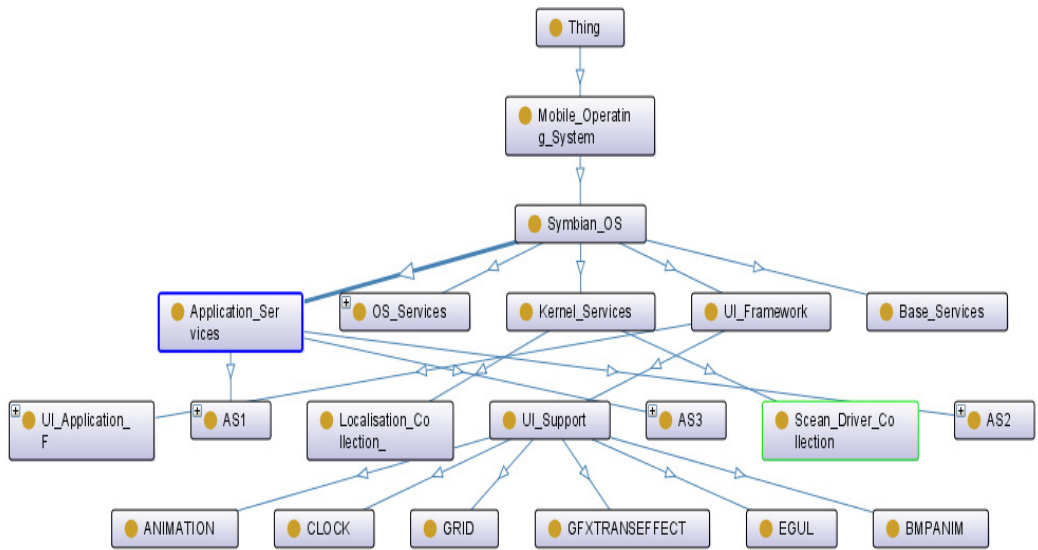

Figure 8. Ontological graph automated by OWL.

### 3.1.2 Android OS

The second tool analyzed during the domain analysis was the Android OS. It is for low powered devices that run on battery, and having plenty of hardware like cameras, light and orientation sensors, Wi-Fi and Universal Mobile Telecommunication Systems (3G telephony) connectivity and a touch-screen. Unlike on other mobile operating systems, Android applications are written in java and run in virtual machines.

The Android OS feature model was developed by identifying essential features from the Android OS user guide [6], Analysis of the Android Architecture Book [7] and by actual day-to-day use of the suite. Figure 9 below illustrate excerpts (condensed feature diagram) from the overall feature model for Android OS.

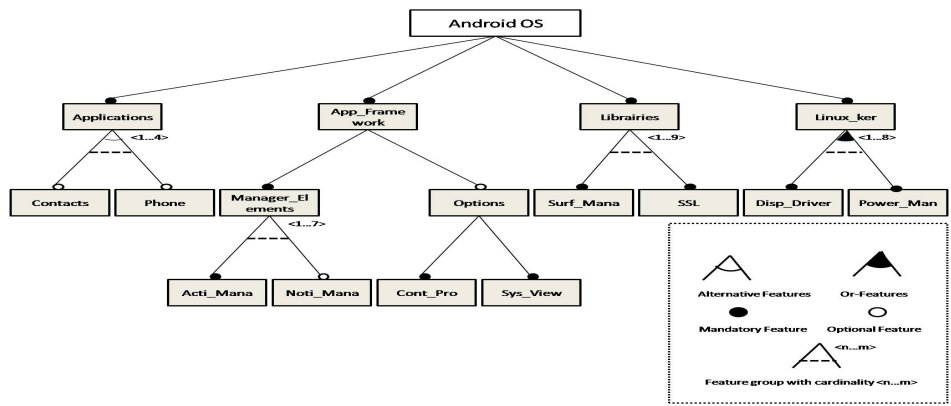

Figure 9. Condensed Android OS feature diagram.

International Journal of Wireless & Mobile Networks (IJWMN) Vol. 4, No. 3, June 2012

From the complete feature model of Android OS, it was possible to extract relevant features (with their descriptions). Each of these features then becomes a candidate for possible inclusion in the federation Ontology. The complete list of Android OS features is listed below in figure 10.

This feature list is taken directly from the Android OS feature model, and generated by the OWL Protégé 4.1 as Ontology classes. As in the case of the Symbian features list, the features aligned as first level at the left of the figure are high-level "parent" features, while those second, third level,... etc represent more detailed "atomic" features.

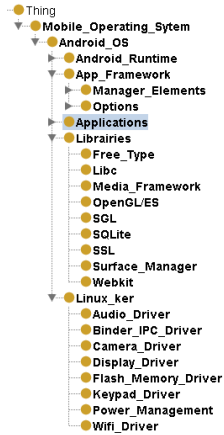

Figure 10. List of Android features generated by the Protégé 4.1.

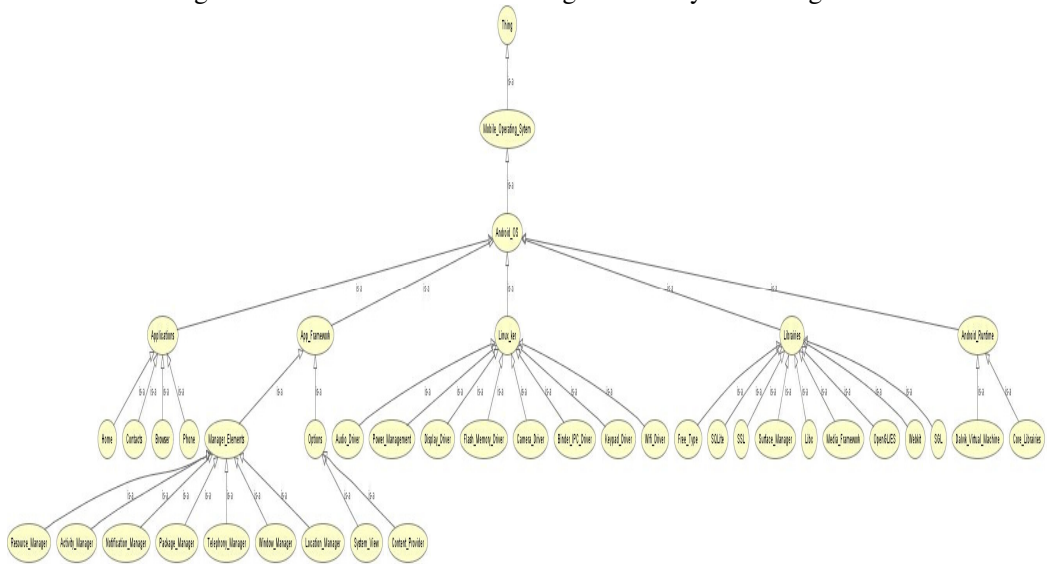

Figure 11. Android visualization Ontology generated by the Protégé 4.1 tool.

Figure 11 is taken from the Ontology editor tool. It gives a thorough idea about the Ontological concepts' hierarchy of android operating system. Meanwhile, figure 12 presents the Ontological graph automated by OWL as layers of classes given by a top down approach.

International Journal of Wireless & Mobile Networks (IJWMN) Vol. 4, No. 3, June 2012

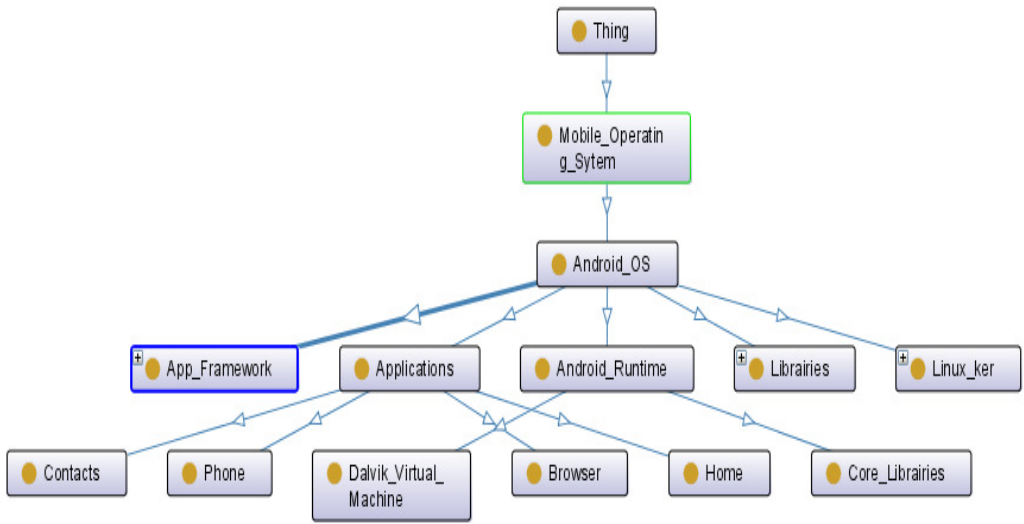

Figure 12. Android Ontological graph automated by OWL.

## 3.2 Federation Ontology

After performing a domain analysis using an in-depth investigation of Symbian OS and Android OS, the two lists were considered together to identify commonalities -- commonalities that would also likely be common with other mobile phone's operating systems--. These commonalities begin to form the list of terms that eventually will make up the federation Ontology.

After identifying these common terms in the domain of mobile phone's operating systems, the words were organized into logical groupings using an "affinity diagram" technique (recall Figure 2). From these affinity diagrams, it was then straight-forward to establish the hierarchical structure of the federation Ontology. The completed federation Ontology classes are shown below in Figure 13.

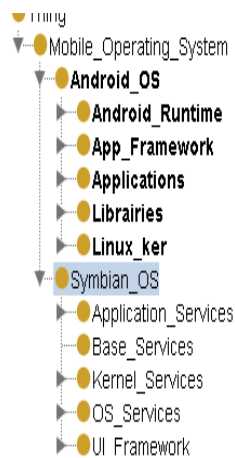

Figure 13. Federation Ontology classes.

### 3.3 Ontology Inter-relationships

The relationships between the three Ontologies were identified and annotated. This was done using UML. Both a top down and bottom up approach were taken to identify the relationships between the three Ontologies and record those relationships in static class diagrams.
Figure 14 is just one example of how the relationships between the three Ontologies are related.

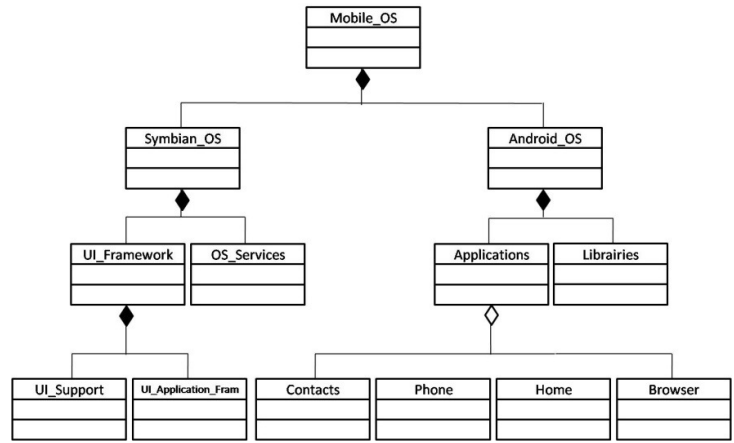

Figure 14. UML representation of the relationships between the three Ontologies.

## 4. CONTRIBUTIONS

While the most important original contribution to the field of wireless cellular communication is to build a mobile phone's operating system Ontology, there are several other contributions:
1. The application of the HFCC to a sample set of mobile phone's operating system (i.e., Symbian and Android) increases the interoperability between the selected set.
2. The methodology was adapted from other sources (notably [3]), but was tailored for identifying and capturing the unique characteristics of mobile phone's operating systems. This methodology can be used to add other mobile phone's operating systems to the tool Ontology. The methodology is as important as the Ontology itself.
3. Three separate Ontologies (and their inter-relationships) are presented: a high-level mobile phone's operating system Ontology, an Ontology that describes the Common Object Model architecture of Symbian, and an Ontology that describes important (from an interoperability viewpoint) classes from Android OS architecture.
4. The mobile operating system Ontology is a modular Ontology. This division into modules has two major advantages: firstly, it facilitates the future introduction of new domain Ontologies, and secondly, it makes the domain (mobile operating system) Ontology more reusable.
5. The main purpose for developing Ontology is to overcome some of the obstacles (such as the limitation of interoperability, lack of communication, and poor shared understanding.
6. The use of feature models as a key asset to manage the commonalities and the variabilities of the mobile phone's operating systems.

## 5. CONCLUSION

This article presented the methodology of a research effort devoted to establishing the set of mobile phone's operating systems Ontologies for integration into the HFCC.  It summarizes the results of the domain analysis undertaken to produce the federation Ontology. Besides, it presents the details of the federation Ontology as well as the two specific tool Ontologies. Finally, the article presents the results of how the three Ontologies inter-relate by using UML to annotate the inter-relationships.

While this methodology and strategy was useful in constructing the feature tree for Symbian and Android, it is only provide a guide for the actual work.

During this Ontology construction process, [8]'s guidelines for Ontology construction were adhered to as much as possible: clarity, coherence, extensibility, minimal Ontological commitment, and minimal encoding bias. However, because it was necessary to adhere closely to the actual class constructs of the tools themselves, it was often not possible to satisfy each of these guidelines.

The mobile phone's operating systems federation Ontology is then left open for further future Enhancement and extension.